\documentclass[aps,prb,onecolumn,showpacs,floatfix,superscriptaddress]{revtex4-2}
\usepackage{amsmath, amssymb, amsxtra, latexsym, amscd, amsthm}
\usepackage[mathscr]{eucal}
\usepackage{mathtools}  % Extends amsmath with additional features
\usepackage{siunitx}

\usepackage{graphicx}
\usepackage{xcolor}
\usepackage{tikz}
\usepackage{svg}

\usepackage{titlesec}   % Customizes section title format
\usepackage{mdframed}% Framed environments with optional colors
\usepackage{subcaption}

\usepackage{natbib}
\newmdenv[linecolor=black,skipabove=\topsep,skipbelow=\topsep,
leftmargin=-10pt,rightmargin=-10pt,
innerleftmargin=10pt,innerrightmargin=10pt]{mybox}

\begin{document}
	\title{Analysis of nonlinear dynamics in a single magnetic pendulum driven by plate-magnet interaction}
	
	\author{Hoang X. Nguyen}
	\affiliation{Hanoi-Amsterdam High School for the Gifted, Hanoi 10000, Vietnam}
	\email{nxhoang22@gmail.com}
	
	\author{Duy V. Nguyen}
	\affiliation{Faculty of Computer Science, Phenikaa University, Hanoi 12116, Vietnam}
	\affiliation{Phenikaa Institute for Advanced Study, Phenikaa University, Hanoi 12116, Vietnam}
	\email{duy.nguyenvan@phenikaa-uni.edu.vn}
	
	\begin{abstract}
		\setlength{\parindent}{0pt}
		
		We analyzed theoretically the nonlinear dynamics of a strong magnetic pendulum consisting of a cylindrical neodymium magnet swinging into a metal plane. The heavy damping of oscillations of the pendulum is caused by eddy currents induced in the metal plate. In this paper, we proposed a mathematical model to describe the braking force acting on the magnetic pendulum and its nonlinear motion.
	\end{abstract}
	\maketitle
	
	\section{Introduction}
	\setlength{\parindent}{0pt}
	
	The subject of magnetic braking has indeed attracted attention and has been discussed in numerous papers. These have included experiments with magnetic braking of a falling magnet inside a non-magnetic conductive tube \cite{Donoso_2009,Donoso2011,Irvine2014}, Thomson jumping \cite{10.1119/1.19407,10.1119/1.1376377} and homopolar motor \cite{Wong2009,Wong200902,Stewart2006,Do2023}. These experiments are frequently shown in Physics lecture demonstrations and in open-day science shows. In recent years, several papers have described experiments with a magnet pendulum for use in Physics teaching and learning \cite{10.1119/1.5028247, Kraftmakher_2007,10.1119/1.4725416,WIJATA2021107229}. The magnetic braking force causes the oscillations to be damped or to transition into chaotic motion. A recent YouTube video showcases a magnetic pendulum consisting of a cylindrical neodymium magnet released towards a copper plate placed at pendulum equilibrium position. When this magnet is released from its equilibrium position and moved towards the copper plate, its oscillation is quickly damped and it stops right in front of the plate \cite[5:06-5:47]{youtube2016}. In this paper, we go beyond a qualitative discussion of the dynamics of the magnetic pendulum suspended by eddy currents induced in a copper plane that is placed perpendicularly to the oscillation plane of the pendulum. We also present a theory that quantitatively accounts for the experimental observations. The theory is accessible to students with only an intermediate understanding of Maxwell's equations in their differential form  and basic knowledge of numerical techniques for solving differential equations.
	
	\section{Theory}
	\subsection{Magnetic braking force}
	\setlength{\parindent}{0pt}
	
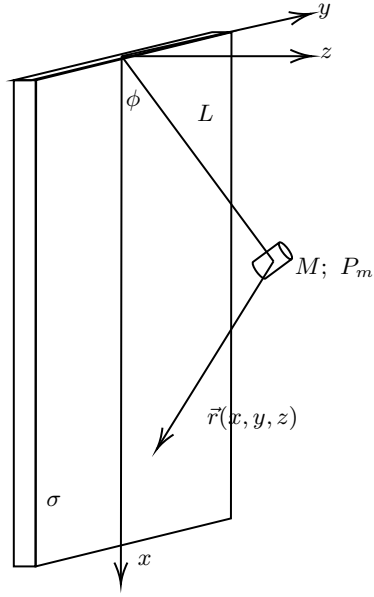
\begin{figure}[h!]
		\centering
		\tikzset{every picture/.style={line width=0.75pt}} %set default line width to 0.75pt        
		
		\begin{tikzpicture}[x=0.75pt,y=0.75pt,yscale=-1,xscale=1]
			
			%Straight Lines [id:da07939696778706895] 
			\draw    (137.39,26.88) -- (214,130.4) ;
			%Shape: Can [id:dp08379089140767881] 
			\draw   (223.11,129.29) -- (209.93,139.06) .. controls (209.24,139.57) and (207.28,138.11) .. (205.57,135.8) .. controls (203.85,133.48) and (203.02,131.19) .. (203.72,130.67) -- (216.9,120.91) .. controls (217.59,120.39) and (219.55,121.85) .. (221.26,124.17) .. controls (222.98,126.48) and (223.81,128.78) .. (223.11,129.29) .. controls (222.42,129.81) and (220.46,128.35) .. (218.75,126.03) .. controls (217.03,123.72) and (216.2,121.42) .. (216.9,120.91) ;
			%Straight Lines [id:da2669064855409651] 
			\draw    (137.39,26.88) -- (137,291.4) ;
			\draw [shift={(137,293.4)}, rotate = 270.08] [color={rgb, 255:red, 0; green, 0; blue, 0 }  ][line width=0.75]    (10.93,-3.29) .. controls (6.95,-1.4) and (3.31,-0.3) .. (0,0) .. controls (3.31,0.3) and (6.95,1.4) .. (10.93,3.29)   ;
			%Straight Lines [id:da3577153042142578] 
			\draw    (137.39,26.88) -- (231.05,5.84) ;
			\draw [shift={(233,5.4)}, rotate = 167.34] [color={rgb, 255:red, 0; green, 0; blue, 0 }  ][line width=0.75]    (10.93,-3.29) .. controls (6.95,-1.4) and (3.31,-0.3) .. (0,0) .. controls (3.31,0.3) and (6.95,1.4) .. (10.93,3.29)   ;
			%Straight Lines [id:da09813161206915777] 
			\draw    (137.39,26.88) -- (231.39,26.88) ;
			\draw [shift={(233.39,26.88)}, rotate = 180] [color={rgb, 255:red, 0; green, 0; blue, 0 }  ][line width=0.75]    (10.93,-3.29) .. controls (6.95,-1.4) and (3.31,-0.3) .. (0,0) .. controls (3.31,0.3) and (6.95,1.4) .. (10.93,3.29)   ;
			%Straight Lines [id:da21471224358840102] 
			\draw    (214,130.4) -- (156.06,223.7) ;
			\draw [shift={(155,225.4)}, rotate = 301.84] [color={rgb, 255:red, 0; green, 0; blue, 0 }  ][line width=0.75]    (10.93,-3.29) .. controls (6.95,-1.4) and (3.31,-0.3) .. (0,0) .. controls (3.31,0.3) and (6.95,1.4) .. (10.93,3.29)   ;
			%Shape: Rectangle [id:dp08250519446152116] 
			\draw   (83,39) -- (94,39) -- (94,284.4) -- (83,284.4) -- cycle ;
			%Shape: Parallelogram [id:dp7768894559951787] 
			\draw   (94.27,38.97) -- (192.78,14.76) -- (192.51,260.18) -- (94,284.4) -- cycle ;
			%Shape: Parallelogram [id:dp40721670622995854] 
			\draw   (183,14.76) -- (191.78,14.76) -- (91.78,39) -- (83,39) -- cycle ;
			
			% Text Node
			\draw (223.26,127.17) node [anchor=north west][inner sep=0.75pt]   [align=left] {$\displaystyle M;\ P_{m}$};
			% Text Node
			\draw (174,50) node [anchor=north west][inner sep=0.75pt]   [align=left] {$\displaystyle L$};
			% Text Node
			\draw (98,247) node [anchor=north west][inner sep=0.75pt]   [align=left] {$\displaystyle \sigma $};
			% Text Node
			\draw (144,277) node [anchor=north west][inner sep=0.75pt]   [align=left] {$\displaystyle x$};
			% Text Node
			\draw (235,-1) node [anchor=north west][inner sep=0.75pt]   [align=left] {$\displaystyle y$};
			% Text Node
			\draw (236,21) node [anchor=north west][inner sep=0.75pt]   [align=left] {$\displaystyle z$};
			% Text Node
			\draw (138,42) node [anchor=north west][inner sep=0.75pt]   [align=left] {$\displaystyle \phi $};
			% Text Node
			\draw (179,202) node [anchor=north west][inner sep=0.75pt]   [align=left] {$\displaystyle \vec{r}( x,y,z)$};
			
		\end{tikzpicture}
		\caption{The system described above: A pendulum swinging towards a metal plate.}
		\label{fig:fig_1}
	\end{figure}

	Consider our system consisting of a magnet that we considered as a pure dipole with magnetic moment $P_m$ and weight $M$ connected to a string with length $L$ in interaction with a semi-infinite plate with conductivity $\sigma$ and negligible thickness (FIG. \ref{fig:fig_1}.).The movement is confined in the $Oxz$ plane and the pendulum has an initial angle of $\phi = \phi(0)$ with no initial velocity. The coordinates of the system originates at the center top of the plate, the $Ox$ coordinate is parallel to the plate, and the $Oz$ coordinate is perpendicular to the plate. Because the plate is semi-infinite, the parameters of the $x$, $y$, and $z$ coordinates are: $0 \leq x \leq \infty; -\infty \leq y \leq \infty; 0 \geq z \geq-\delta, \delta \ll L$. \\
	
	To explicitly calculate the magnetic damping force, we will have to firstly go through calculating the eddy currents in the plate from equations (\ref{eq:eq2}) to (\ref{eq:eq9}). Then, the force shall be calculated from equations (\ref{eq:eq10}) to (\ref{eq:eq19}). \\
	
	Moreover, to make the equations less cumbersome, we shall denote short connotations of some terms we will encounter throughout the problem:
	\begin{align}
		\label{eq:eq1}
		&x-L \cos \phi=X; y=Y; z-L \sin \phi=Z \notag \\ 
		&r^2=(x-L \cos \phi)^2+y^2+(z-L \sin \phi)^2 \\
		&\vec{r}=(x-L\cos\phi)\hat{x}+\vec{y}+(z-L\sin \phi)\hat{z} \notag
	\end{align}
	
	The magnet as it swings will induce an electric field due to Faraday's law:
	\begin{equation}
		\label{eq:eq2}
		\vec {\nabla} \times \vec{E} = -\frac{\partial\vec{B}}{\partial t}
	\end{equation}
	
	Rewriting (\ref{eq:eq2}) using the magnetic vector potential $\vec{A}$ through substituting $\vec{B} = \vec{\nabla} \times \vec{A}$ and removing the del operator($\vec{\nabla}$), we will have the equation as in \cite[p.437]{Griffiths2013}:
	\begin{equation}
		\label{eq:eq3}
		\vec{E}= -\vec{\nabla} V -\frac{\partial \vec{A}}{\partial  t} 
	\end{equation}
	
	The gradient of a scalar in (\ref{eq:eq3}) works as the electrostatic potential when $\vec{A}$ is constant \cite[p.437]{Griffiths2013}. However, in this problem we can simply omit the gradient term due to the strength of the magnetic field in the regions close to the magnet makes the effect of the gradient negligible \cite{10.1119/1.4725416}.
	\begin{equation}
		\label{eq:eq4}
		\vec{E} \approx -\frac{\partial \vec{A}}{\partial t}
	\end{equation}
	
	The magnetic vector potential of a dipole can be written as in \cite[p.253]{Griffiths2013}:
	\begin{equation}
		\label{eq:eq5}
		\vec{A}(r)=k_m \frac{\vec{P}_m \times \vec{r}}{r^3}
	\end{equation}
	With $\vec{r}$ is the vector points from the center (source point) of the dipole to the field point (plate) and $k_m=\displaystyle \frac{\mu_0}{4 \pi}$.
	
	Also, the dipole's orientation can be described through unit vectors and rewritten into:
	\begin{equation}
		\label{eq:eq6}
		\vec{P}_m=P_m(-\cos \phi \hat{z}+\sin \phi \hat{x})
	\end{equation}
	
	From (\ref{eq:eq1}), (\ref{eq:eq5}) and (\ref{eq:eq6}), we can split $\vec{A}$ into different components:
	\begin{align}
		\label{eq:eq7}
		&A_x = \frac{k_m P_m y \cos \phi}{r^3} \notag \\
		&{A_y} = \displaystyle - \frac{{{k_m}{P_m}} \left[ {(z - L\sin \phi )\sin \phi  + (x - L\cos \phi )\cos \phi } \right] }{{{r^3}}} \\
		&A_z = \frac{k_m P_m y \sin \phi}{r^3} \notag 
	\end{align}
	
	From (\ref{eq:eq4}) and (\ref{eq:eq7}), we can calculate explicitly $\vec{E}$ running inside the metal plate:
	\begin{align}
		\label{eq:eq8}
		&E_x=k_m P_m \dot{\phi} \frac{Y\left[\sin \phi\left(X^2+Y^2+Z^2\right)+3 L \cos \phi(X \sin \phi-Z \cos \phi)\right]}{(X^2+Y^2+Z^2)^\frac{5}{2}} \notag \\
		&E_y=k_m P_m \dot{\phi} \frac{-(X \sin \phi-Z \cos \phi)\left[X(X+3 L \cos \phi)+Y^2+Z(Z+3 L \sin \phi)\right]}{(X^2+Y^2+Z^2)^\frac{5}{2}}\\
		&E_Z=k_m P_m \dot{\phi} \frac{Y\left[-\cos \phi\left(X^2+Y^2+Z^2\right)+3 L \sin \phi(X \sin \phi-Z \cos \phi)\right]}{(X^2+Y^2+Z^2)^\frac{5}{2}} \notag
	\end{align}
	
	The eddy current's strength is given by Ohm's law in differential form:
	\begin{equation}
		\label{eq:eq9}
		\vec{j}=\sigma \vec{E}
	\end{equation}
	
	To calculate the force interacting with the dipole comes from eddy currents running inside the plate, we can use the Bio-Savart law as the frequency of the pendulum is sufficiently small to consider the time variation of currents in the plate \cite[p.319]{Griffiths2013}. \\
	
	The magnetic field on the magnet caused by eddy currents in the metal plate calculated by the Bio-Savart law:
	\begin{equation}
		\label{eq:eq10}
		\vec{B}(\vec{r}')=k_m \int \frac{\vec{j} \times \vec{r} '}{r^{\prime 3}} d V
	\end{equation}
	With $r^{\prime}=r ; \vec{r}'=-\vec{r} ; d V=d x d y d z$\\
	Rewriting the secondary magnetic field with $\vec{r}$ and $r$, we have:
	\begin{align}
		\label{eq:eq11}
		\vec{B}(r)=-k_m \int \frac{\vec{j} \times \vec{r}}{r^3} d V
	\end{align}
	
	From (\ref{eq:eq8}), (\ref{eq:eq9}) and (\ref{eq:eq11}), we have the components of the secondary magnetic field cause by eddy currents:
	\begin{align}
		\label{eq:eq12}
		&B_x(r)=-k_m \int_V \frac{j_y(z-L \sin \phi)-j_z y}{ \displaystyle \left[(x-L \cos \phi)^2+y^2+(z-L \sin \phi)^2\right]^{\frac{3}{2}}} d V \notag \\
		&B_y(r)=-k_m \int_V \frac{j_z(x-L \cos \phi)-j_x(z-L \sin \phi)}{\displaystyle \left[(x-L \cos \phi)^2+y^2+(z-L \sin \phi)^2\right]^{\frac{3}{2}}} d V \\
		&B_z(r)=-k_m \int_V \frac{j_x y-j_y(x-L \cos \phi)}{\displaystyle \left[(x-L \cos \phi)^2+y^2+(z-L \sin \phi)^2\right]^{\frac{3}{2}}} d V \notag
	\end{align}
	
	The interacting energy of the magnet and the currents in the plate:
	\begin{align}
		\label{eq:eq13}
		W&=-\frac{1}{2} \vec{P}_m \cdot \vec{B} \notag \\
		&=-\frac{1}{2}\left(P_m \sin \phi \hat{x}-P_m \cos \phi \hat{z}\right)\left(B_x \hat{x}+B_z \hat{z}\right) \\
		&=-\frac{1}{2}P_m\left(B_x \sin \phi-B_z \cos \phi\right) \notag
	\end{align}
	
	The force in interaction with the magnet:
	\begin{align}
		\label{eq:eq14}
		\vec{F} = -\vec{\nabla} W
	\end{align}
	
	We will only calculate the projected force that is perpendicular to the string, as the equation of motion of the pendulum only stays in the $Oxz$ plane, and the force tangent to the string is balanced by the tension in the string:\\
	
	The damping force in the tangential direction of the pendulum, $F_\phi$, is defined as follows:
	\begin{align}
		\label{eq:eq15}
		F_\phi=-\frac{1}{L} \frac{\partial W}{\partial \phi}=\frac{P_m}{2L} \frac{\partial}{\partial \phi}\left(B_x \sin \phi-B_z \cos \phi\right)=-\frac{\sigma k_m^2 P_m^2}{2L} \eta_(\phi) \dot{\phi}
	\end{align}
	
	where:
	\begin{align}
		\label{eq:eq16}
		\eta(\phi) =\frac{\partial}{\partial \phi} \int_V \Psi(X, Y, Z, \phi) d X d Y d Z
	\end{align}
	
	with:
	\begin{align}
		\label{eq:eq17}
		\Psi (X,Y,Z,\phi ) = \frac{{ - (X\sin \phi  - Z\cos \phi )\left\{ {3L{Y^2} + (X\cos \phi  + Z\sin \phi )\left[ {{X^2} + {Y^2} + {Z^2} + 3L(X\cos \phi  + Z\sin \phi )} \right]} \right\}}}{{{{\left( {{X^2} + {Y^2} + {Z^2}} \right)}^4}}}
	\end{align}
	
	Taking the derivative of $\Psi(X,Y,Z, \phi)$ with respect to $\phi$ in (\ref{eq:eq17}), we have:
	\begin{align}
		\label{eq:eq18}
		\eta(\phi)= \int_V \Gamma(x, y, z, \phi)dxdydz
	\end{align}
	
	with:
	\begin{align}
		\label{eq:eq19}
		\Gamma(x, y, z, \phi)=&\frac{24 y^2 L^2(x \sin \phi-z \cos \phi)^2}{\left[(x-L \cos \phi)^2+y^2+(z-L \sin \phi)^2\right]^5}-\frac{3 y^2 L(x \cos \phi+z \sin \phi)}{\left[(x-L \cos \phi)^2+y^2+(z-L \sin \phi)^2\right]^4} \\
		&+\frac{-(x \cos \phi+z \sin \phi)(x \cos \phi+z \sin \phi-L)}{\left[(x-L \cos \phi)^2+y^2+(z-L \sin \phi)^2\right]^3}+\frac{(x \sin \phi-z \cos \phi)^2}{\left[(x-L \cos \phi)^2+y^2+(z-L \sin \phi)^2\right]^3} \notag \\
		&+\frac{6 L(x \cos \phi+z \sin \phi-L)(x \sin \phi-z \cos \phi)^2}{\left[(x-L \cos \phi)^2+y^2+(z-L \sin \phi)^2\right]^4}+\frac{-3 L(x \cos \phi+z \sin \phi)(x \cos \phi+z \sin \phi-L)^2}{\left[(x-L \cos \phi)^2+y^2+(z-L \sin \phi)^2\right]^4} \notag \\
		&+\frac{6 L(x \sin \phi-z \cos \phi)^2(x \cos \phi+z \sin \phi-L)}{\left[(x-L \cos \phi)^2+y^2+(z-L \sin \phi)^2\right]^4}+\frac{24 L^2(x \sin \phi-z \cos \phi)^2(x \cos \phi+z \sin \phi-L)^2}{\left[(x-L \cos \phi)^2+y^2+(z-L \sin \phi)^2\right]^5}\notag
	\end{align}
	
	Using an integral calculator and inserting \(L=\SI{1}{\meter} \), we can draw the dependencies of $\eta(\phi)$ to $\phi$ in (FIG. \ref{fig:fig_2}.):
	\begin{figure}[h!]
		\centering
		\includegraphics[scale=0.4]{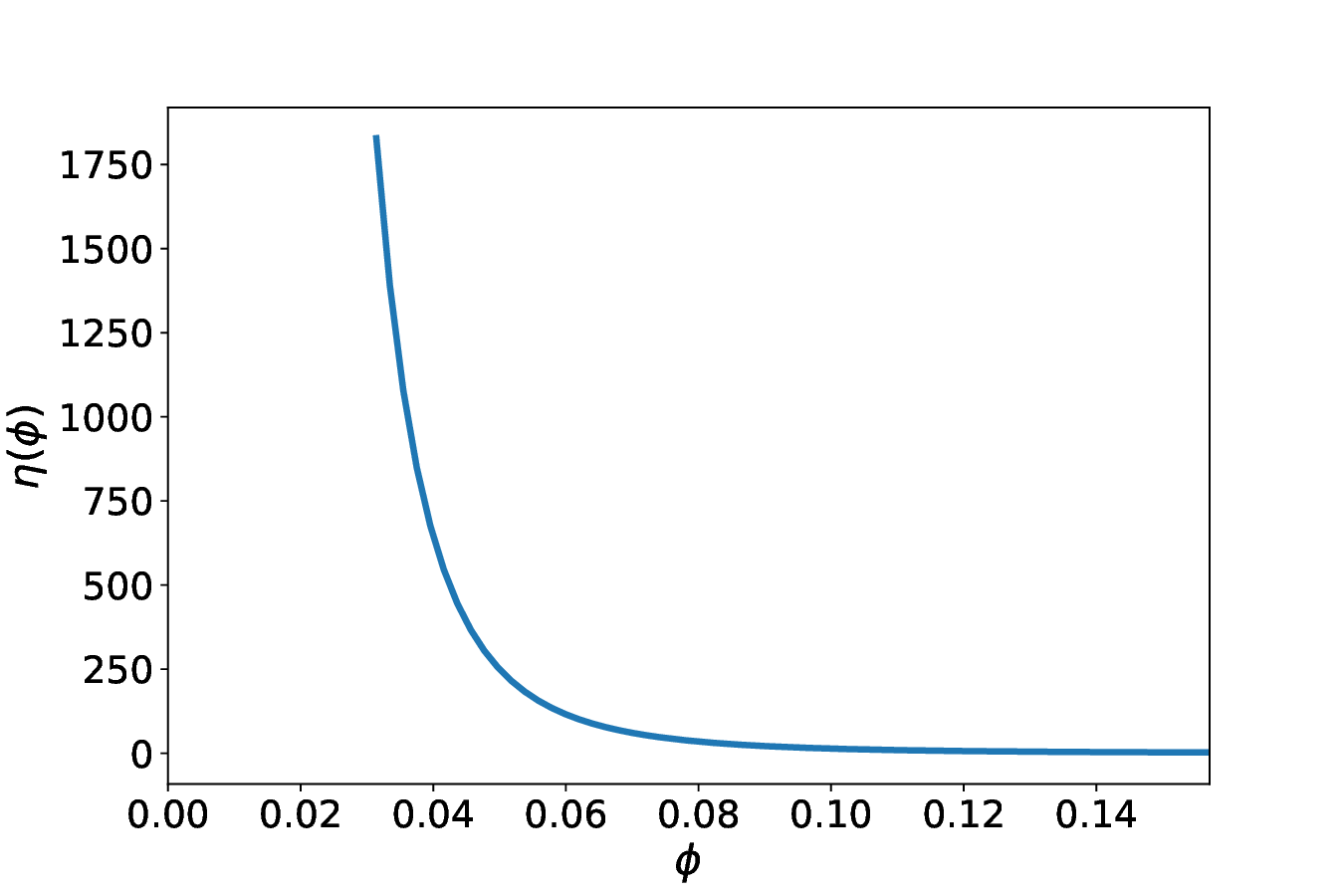}
		\caption{Dependence of $\eta(\phi)$ to $\phi$.}
		\label{fig:fig_2}
	\end{figure}
	
	From (FIG. \ref{fig:fig_2}.), we can easily see that $\eta(\phi)$ would reach to a horizontal asymptote very rapidly from a vertical asymptote at $\phi = 0.02$ when $\phi$ starts to reach $0.05$. In our theoretical model, it shows that $\eta(\phi)$ will not play as much of an important role in the motion of the pendulum while $\phi$ is large than when $\phi \approx 0$. \\
	
	\subsection{Dynamics of the magnetic pendulum}
	\setlength{\parindent}{0pt}
	
	The equation of motion projected perpendicular to the string:
	\begin{equation}
		\label{eq:eq20}
		M L \ddot{\phi}=-M g \sin \phi+F_\phi\\
	\end{equation}
	With $F_\phi$ the damping force on the magnet projected perpendicular to the string found in (\ref{eq:eq15}).
	
	Substituting (\ref{eq:eq15}) into (\ref{eq:eq20}) and simplifying it, we have our DE:
	\begin{equation}
		\label{eq:eq21}
		\ddot{\phi}+\frac{\sigma k_m^2 P_m^2}{2ML^2} \eta_(\phi) \dot{\phi}+\frac{g}{L} \sin \phi=0
	\end{equation}
	
	Now, we will substitute the numerical values of the magnetic moment of the dipole as \(P_m=\SI{1.29}{\ampere\meter\squared}\) \cite{Amrani2015} and the mass of the magnet is \(M=\SI{10}{\gram}\). The conductivity of the metal surface (copper)  \(\sigma=\SI{5.85e8}{\siemens\per\meter}\) \cite{Chapman1999} and  thickness of copper plate \(\delta=\SI{1e-3}{\meter}\), and the gravitational acceleration is \(g=\SI{9.8}{\meter\per\second\squared}\). The initial angle and angular velocity of the magnet is \(\phi(0)=\SI{0.33}{\radian}\) and \(\dot{\phi}(0)=\SI{0}{}\) respectively. \\
	
	Inserting (\ref{eq:eq21}) into the computer and using the Runge–Kutta method at the fourth order (RK4) \cite[p.295-333]{johansson2019numerical}, we get the dependence of $\phi(t)$, $\dot{\phi}(t)$ and $F_\phi(t)$ to $t$ as below:
	\begin{figure}[h!]
		\centering
		\begin{subfigure}[t]{0.45\textwidth}
			\centering
			\includegraphics[scale=0.35]{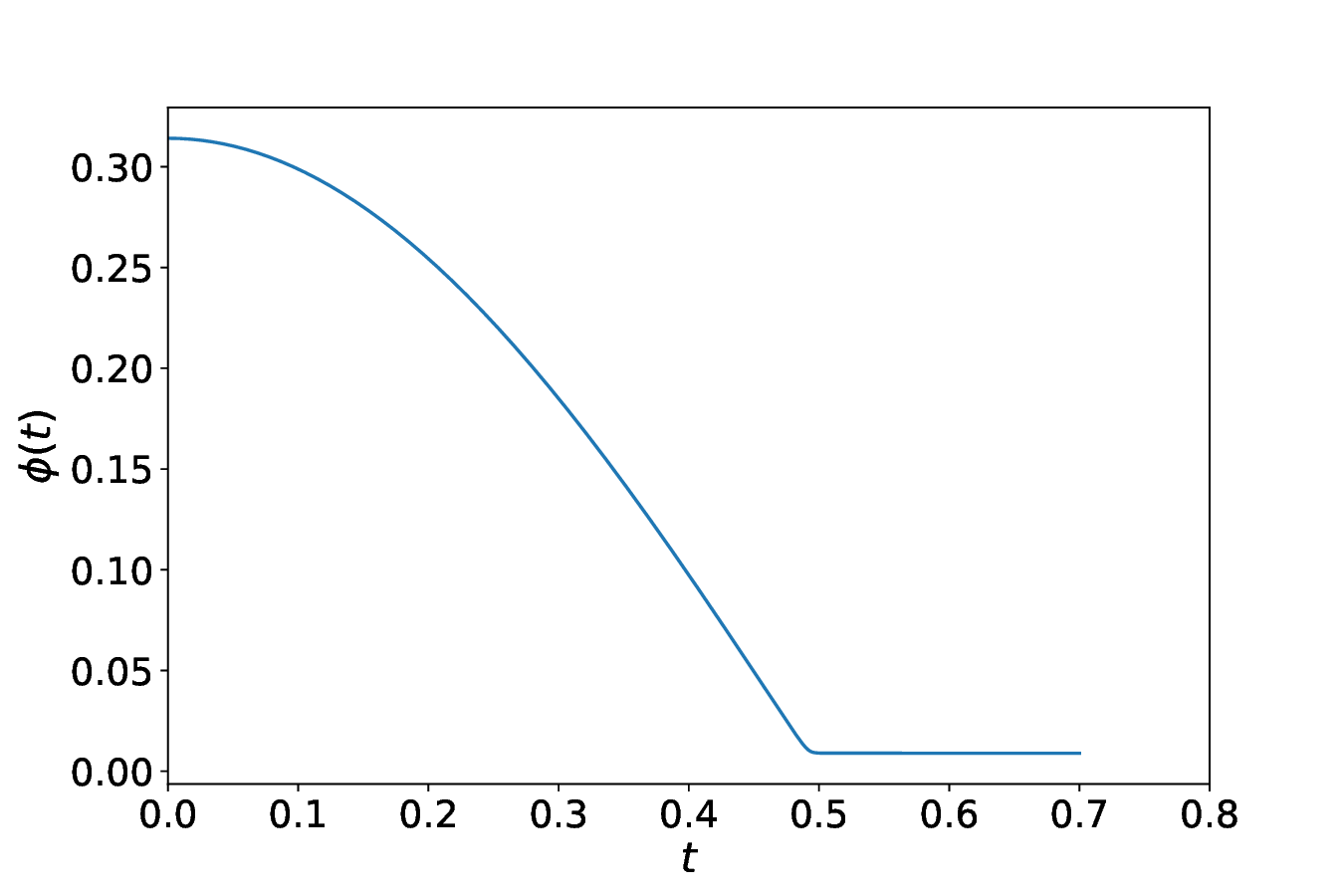}
			\caption{Dependence of $\phi(t)$ on $t$.}
			\label{fig:fig_3a}
		\end{subfigure}
		\hfill
		\begin{subfigure}[t]{0.45\textwidth}
			\centering
			\includegraphics[scale=0.35]{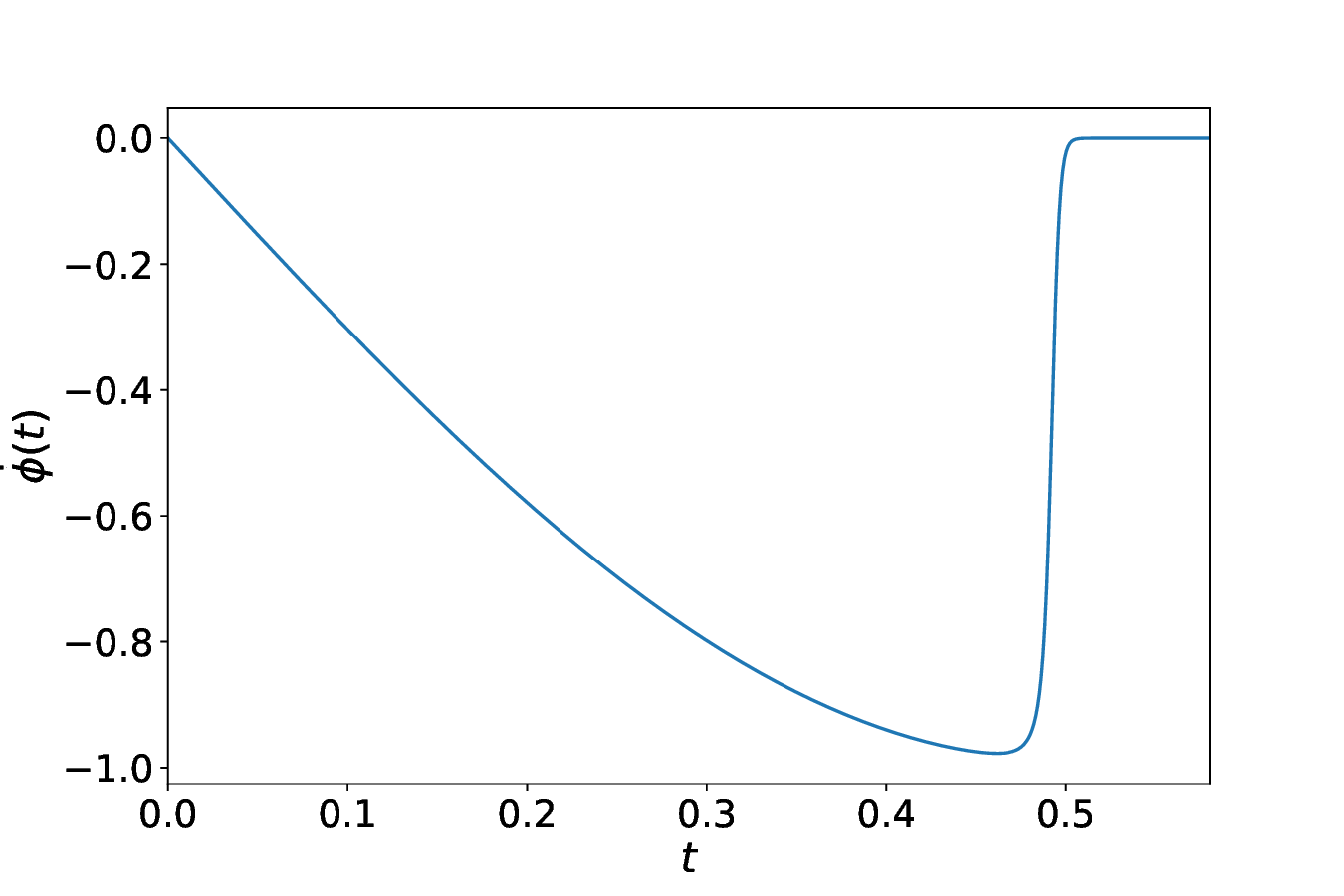}
			\caption{Dependence of $\dot{\phi}(t)$ on $t$.}
			\label{fig:fig_3b}
		\end{subfigure}
		\caption{Comparison of $\phi(t)$ and $\dot{\phi}(t)$ dependencies on time $t$.}
		\includegraphics[scale=0.4]{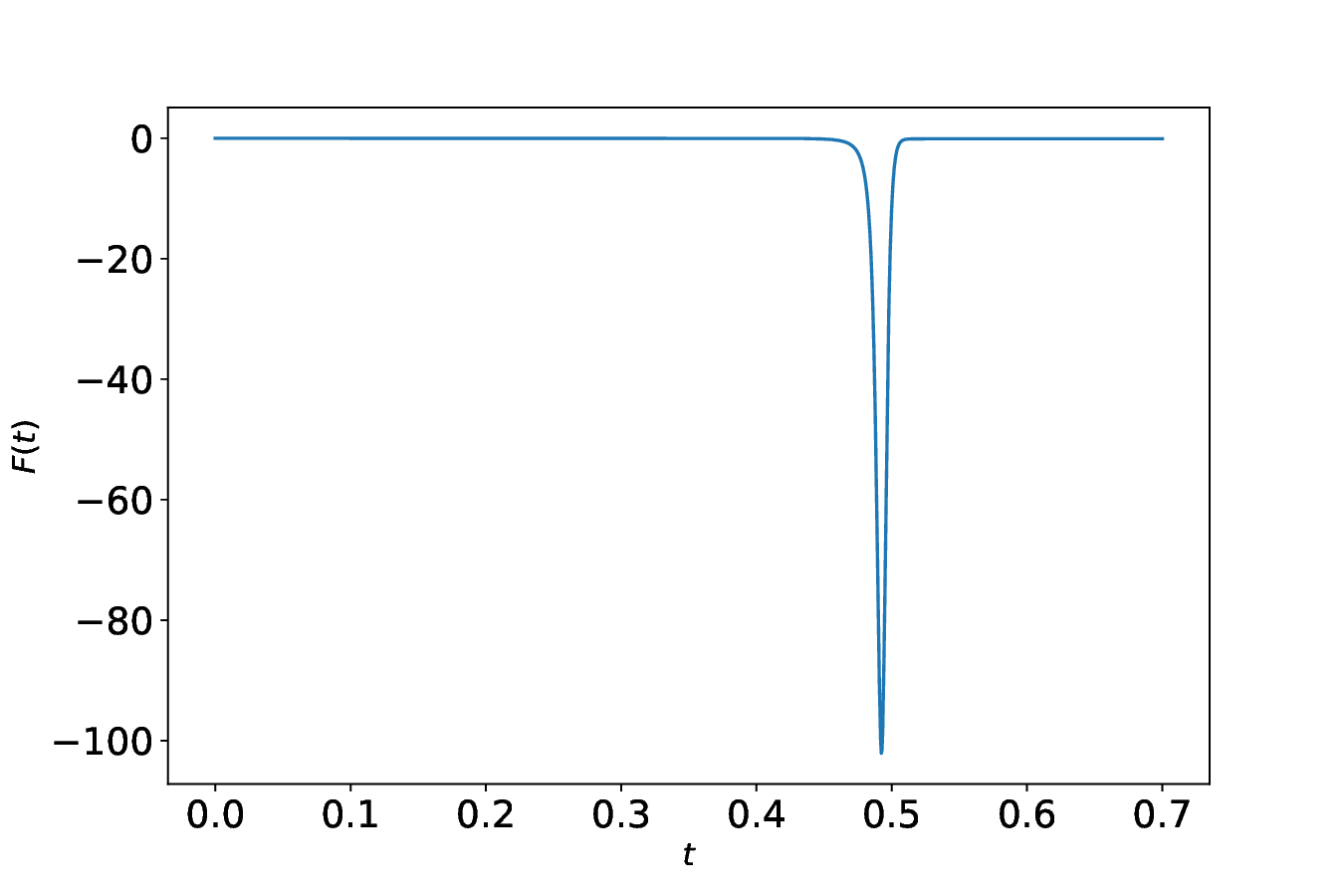}
		\caption{Dependence of $F_\phi(t)$ over $t$.}
		\label{fig:fig_4}
	\end{figure}
	
	From (FIG. \ref{fig:fig_4}), we can clearly see that our calculated results show that the motion of the pendulum can be separated into 2 different periods. The first period is when $\phi=\phi(0)$ to around $0.48$ seconds, where the system can be treated as a normal pendulum ($F_\phi \approx 0$). The second period happens in about $0.1$ seconds after that, agreeing with (FIG. \ref{fig:fig_2}.) that the braking force increases exponentially to halt the motion of the pendulum instantaneously. This also seems to be the phenomena that \cite[5:43-5:47]{youtube2016} depicts, as the magnet in the YouTube video stops immediately when it is very close to the metal plate.
	
	\section{Conclusion}
	\setlength{\parindent}{0pt}
	
	In this paper, we have successfully created a theoretical solution for the motion of a magnetic pendulum interacting with a metal surface. The solution described the velocity and position of the magnet over time, and gives us a qualitative look on the strength of magnetic damping forces. The results above could serve as a valuable teaching tool in classroom experiments, which would provide an example of nonlinear motion in physical systems.
	
	% \section*{Acknowledgments}
	\bibliographystyle{apsrev4-2}
	\bibliography{ref}
	
\end{document}